\documentclass[aps,floats,prl,
showpacs, twocolumn ]{revtex4}

\usepackage{amsfonts,amsmath,mathrsfs} \usepackage{bm} \usepackage{dcolumn}
\usepackage{graphicx}
 \usepackage{latexsym}

\begin{document}

\title{Local diffusion theory of localized waves in open media}

\author{Chu-Shun Tian$^1$, Sai-Kit Cheung$^2$ and Zhao-Qing Zhang$^2$}

\affiliation{$^{1}$ Institut f\"{u}r Theoretische Physik,
Universit\"{a}t zu K\"{o}ln, D-50937 K\"{o}ln, Germany \\
$^{2}$ Department of Physics, Hong Kong University of Science and
Technology, Clear Water Bay, Kowloon, Hong Kong}

\begin{abstract}
{\rm We report a first-principles study of static transport of
localized waves in quasi-one-dimensional open media. We found that
such transport, dominated by disorder-induced resonant
transmissions, displays novel diffusive behavior. Our analytical
predictions are entirely confirmed by numerical simulations. We
showed that the prevailing self-consistent localization theory [van
Tiggelen, {\it et. al.}, Phys. Rev. Lett. \textbf{84}, 4333 (2000)]
is valid only if disorder-induced resonant transmissions are
negligible. Our findings open a new direction in the study of
Anderson localization in open media.}
\end{abstract}


\maketitle

{\it Introduction.}---In the past years experimental studies of
localization have been boosted due to the unprecedented level of
manipulating ultracold atomic gases \cite{Billy08}, dielectric
materials
\cite{Fishman07,Wiersma08,Genack06,Chabanov00,Maret06,Zhang09}, and
elastic media \cite{Hu08}. A key feature shared by many experimental
setups \cite{Maret06,Chabanov00,Zhang09,Chabanov03,Hu08} is that,
there, one allows wave energies to leak out of systems through
boundaries in order to facilitate measurements. Consequently, wave
interferences interplay strongly with the wave energy leakage that,
conceptually, enriches transport phenomena of localized waves while,
technically, pushes forward developments of theoretical approaches.
In particular, as a unique property of finite-sized samples,
localized states in the sample center create resonant transmissions
\cite{Azbel83,Freilikher03}. Although these transmissions are rare
events, nevertheless, they contribute significantly to average
transmission. In fact, random matrix theory predicts that in
quasi-one-dimensions (quasi-$1$D), the localization length measured
from average transmission can be four times larger than that
measured from typical transmission \cite{Beenakker97}. Recently,
disorder-induced resonant transmissions have found considerable
practical applications. For example, they mimic a ``resonator'' with
high-quality factors and thus are used to fabricate random laser
\cite{Cao02} and to realize optical bistability \cite{Freilikher10}.

However, to study this intriguing interplay has proved to be, in
general, a formidable task. Confronting this challenge, a decade ago
van Tiggelen, Lagendijk, and Wiersma took a bold stroke
\cite{Lagendijk00}, hypothesizing a so-called self-consistent local
diffusion (SCLD) model for localization in open media (for a review,
see Ref.~\cite{Vollhardt80}). They phenomenologically generalized
the self-consistent localization theory of infinite media
\cite{Vollhardt80} by demanding the diffusion coefficient to be
position-dependent so as to take boundary effects into account. The
SCLD model (as well as its dynamic generalization
\cite{Hu08,Skipetrov06}), having the advantages of physical
transparency and methodological simplicity over other approaches, is
guiding considerable experimental and theoretical activities (e.g.,
Refs.~\cite{Zhang09,Hu08,Skipetrov09}). Yet,
the validity of the SCLD model inside localized samples is largely
unknown and, in fact, has been severely questioned by recent pulsed
microwave experiments \cite{Zhang09}. There, it was shown that the
(dynamic) SCLD model \cite{Skipetrov06} fails to describe transport
in quasi-$1$D localized samples at long times, when energies are
mainly stored in long-lived modes. The dramatic discrepancy
\cite{Zhang09} between experimental measurements and theoretical
predictions is conveying an opinion. That is, the highly non-local
object of disorder-induced resonant transmission \cite{Azbel83},
which plays a decisive role in transport of localized waves
\cite{Freilikher03}, may not be captured by such a model.

Motivated by these activities, we performed a first-principles study
of static wave transport in quasi-$1$D localized samples, i.e.,
$L\gg \xi$ with $L$ and $\xi$ the sample and localization length,
respectively. We predicted analytically and confirmed numerically
that in these systems, localized waves display a novel diffusion
phenomenon. Our theory shows that the SCLD model is valid only if
disorder-induced resonant transmissions are negligible. Our
findings, capable of being generalized to higher dimensions, may
open a new direction in the study of Anderson localization in open
media.

\begin{figure}[h]
 \centering
 \includegraphics[width=8.0cm]{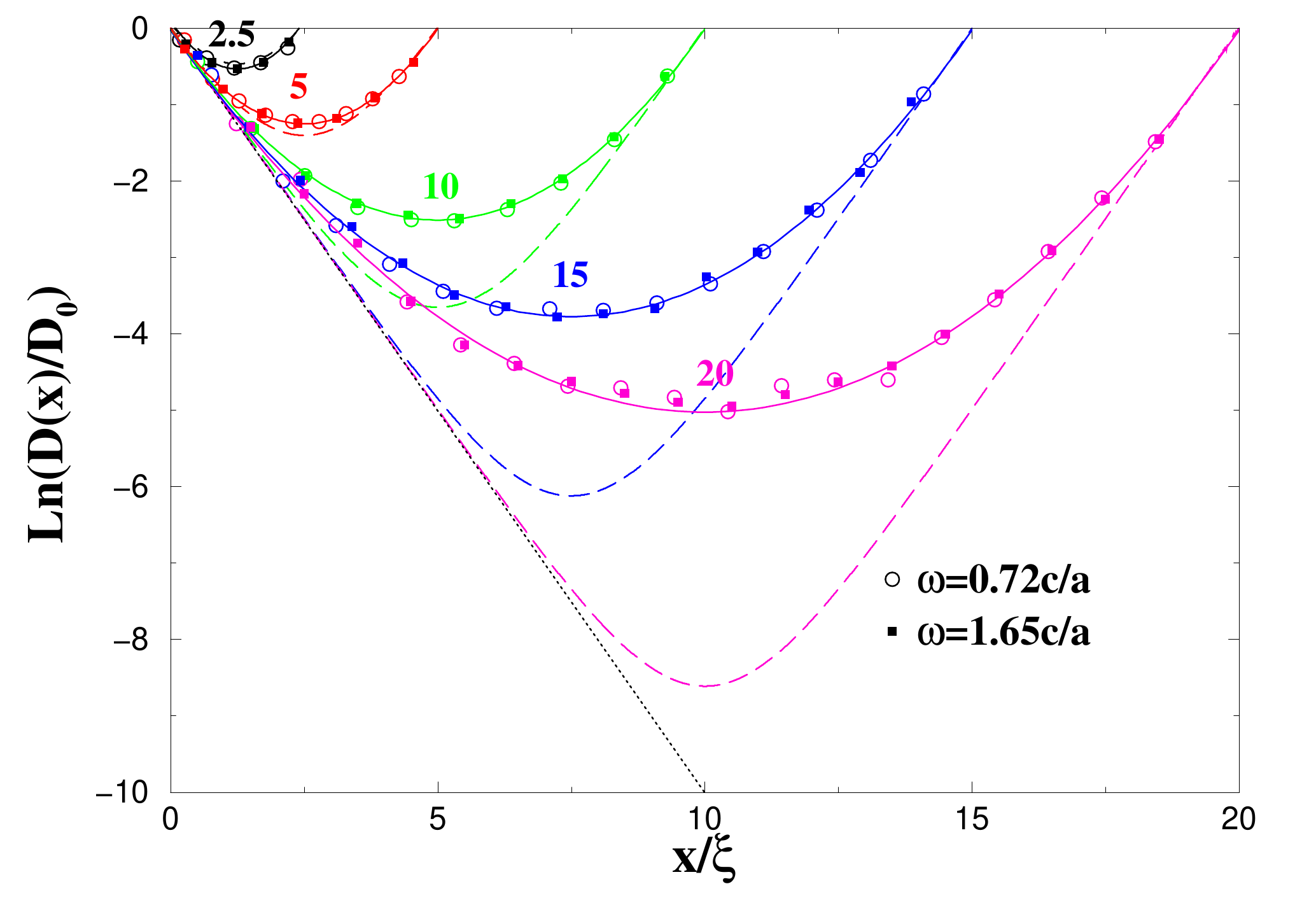}
 \caption{Comparing results obtained from numerical simulations,
 the analytical prediction (\ref{DSL}) (solid lines) and the SCLD model (dashed lines).
 $D(x)/D_0$ was computed numerically for two wave frequencies,
 $\omega=1.65 c/a$ (square) and $\omega=0.72 c/a$
 (circle), and for five different sample lengths, $L/\xi=2.5,5,10,15 $ and $20$\,.}
 \label{1Dlocal}
\end{figure}

{\it Main results and qualitative discussions.}---We considered the
wave intensity and particularly its spatial correlation function,
${\cal Y} (x,x')$\,. Our first-principles analytic theory,
justifying the static local diffusion equation: $-\partial_x
D(x)\partial_x {\cal Y} (x,x') = \delta(x-x')$ with $x$ ($x'$) the
distance of the observation (source) point from given boundary,
leads to the following central results. (i) The local (or
position-dependent) diffusion coefficient $D(x)$ displays a novel
scaling behavior. Specifically, $D(x)$ depends on $x$ via the
scaling $\lambda=(L-x)x/(L\xi)$ [$D_0=D(0)$],
\begin{eqnarray}
D(x)/D_0=D_\infty(\lambda)\,, \label{scalingtransformation}
\end{eqnarray}
and the scaling function $D_\infty(\lambda)$ is $\sim e^{-\lambda}$
for $\lambda \rightarrow \infty$\,. (ii) From (i) it follows that
inside the sample, surprisingly, $D(x)$ is enhanced drastically from
the exponential decay,
\begin{equation}
D(x)/D_0 \propto e^{x^2/(L\xi)}e^{-x/\xi} \,, \qquad \xi \ll x \leq
L/2 \,. \label{DSL}
\end{equation}
[$D(x)=D(L-x)$\,.] (iii) Eqs.~(\ref{scalingtransformation}) and
(\ref{DSL}) are universal regardless of the time-reversal symmetry.
Our results, while entirely confirmed by simulations
(Fig.~\ref{1Dlocal}), show that the SCLD model
fails in localized samples.

Let us first present qualitative explanations of the main results
(i)-(iii). The scaling behavior, Eq.~(\ref{scalingtransformation}),
finds its origin in wave interferences. Indeed, as waves penetrate
into a time-reversal medium, they may counterpropagate along the
same loop and interfere with each other---the well-known weak
localization \cite{Anderson79}. However, different from infinite
media, in open media, wave energies leak out through the boundaries
and as such, the probability (in the frequency domain) of forming a
loop is finite in the static limit, which is $\propto (L-x)x/L$ in
quasi-$1$D. Then, a one-loop wave interference correction to $D_0$
results, which is $x$-dependent and of the order of $\lambda =
(L-x)x/(L\xi)$\,. Furthermore, because $\lambda$ monotonously
decreases from the sample midpoint to boundaries, as waves propagate
towards the sample center, the propagation paths tend to form more
(say $n$) loops with a probability $\propto [(L-x)x/L]^n$\,, leading
to a wave interference correction $\sim \lambda^n$\,. Thus, wave
interferences everywhere render $D(x)$ depending on $x$ via
$\lambda$\,.

That the scaling behavior of local diffusion is far beyond the reach
of the SCLD model is best appreciated by broken time-reversal
systems (unitary symmetry). There, the above one-loop wave
interference is absent and, therefore, the SCLD model ceases to
work. [Indeed, for $\lambda\rightarrow 0$\,, the linear term in the
$\lambda$-expansion of $D_\infty(\lambda)$ now disappears.] Instead,
two paths may take the same route, propagate in the same direction
but visit individual scatterers at different times. As such, they
form loops and equally contribute wave interference corrections to
$D_0$\,. The perturbative $\lambda$-expansion is thereby justified,
with the leading order correction $\sim \lambda^2$\,.

Having explained Eq.~(\ref{scalingtransformation}), let us estimate
the asymptotic form of $D_\infty(\lambda)$ at $\lambda\rightarrow
\infty$\,. To this end we enjoy the universality of
$D_\infty(\lambda)$ and set $L\rightarrow \infty$ (semi-infinite
media). In this simple case, on physical grounds, we expect
$D(x)/D_0=D_\infty(\lambda)\sim e^{-x/\xi}$ for $x\gg \xi$\,.
Because of $\lambda=x/\xi$\,, we find $D_\infty(\lambda)\sim
e^{-\lambda}$ for $\lambda\gg 1$\,. This asymptotic form then gives
Eq.~(\ref{DSL}) for finite-sized samples. Importantly, the
significant enhancement from the exponential decay may be related to
the fluctuation of the inverse localization length $\gamma$ in
finite-sized samples. Indeed, for $L\gg \xi$\,, the distribution of
$\gamma$ is Gaussian, with the average and variance being $\xi^{-1}$
and $2/(\xi L)$\,, respectively \cite{Beenakker97,Anderson80}.
Averaging $e^{-\gamma x}$\,, we obtained
\begin{equation}
\int_0^\infty d\gamma\, e^{-\gamma x} e^{-\frac{\xi
L}{4}\left(\gamma-\xi^{-1}\right)^2} \sim D(x)
\label{Lfluctuations}
\end{equation}
for $x$ deep inside the sample.

{\it Failure of SCLD model.}---For
simplicity we focus on classical scalar waves, and begin with
testing the validity of the SCLD model \cite{Lagendijk00}. We
performed numerical simulations of the spatially-resolved wave
intensity across a randomly layered medium, which is embedded in an
air background and excited by a plane wave of (angular) frequency
$\omega$\,. The layer thickness is $a$\,, and the relative
permittivity at each layer fluctuates independently, with a uniform
distribution in the interval $[1-\sigma,1+\sigma]$\,. Here $\sigma$
measures the degree of randomness of the system and throughout this
work we considered non-reflecting boundaries. We set $\sigma=0.7$
and considered two wave frequencies, $\omega=1.65c/a$ and
$0.72c/a$\,, where $c$ is the speed of light in the air. We used the
standard transfer matrix method to calculate the transmission
coefficient, $T_\beta$\,, and wave intensity distribution,
$I_\beta(x)$\,, for each configuration $\beta$\,. For each $\omega$,
we calculated the ensemble-averaged current $j\equiv \langle
T_\beta\rangle$ and wave intensity distribution $I(x)\equiv \langle
I_\beta(x)\rangle
$ of $2,000,000$ realizations of dielectric disorders for different
sample lengths. Since the current across the sample is uniform due
to the conservation law, we used the relation: $j=-D(x) \partial_x
I(x)$ to compute $D(x)$ by presuming the static local diffusion
equation.

In Fig.~\ref{1Dlocal} the results of $D(x)/D_0$ obtained by
simulations and by numerically solving the SCLD model are presented.
First of all, they both show that $D(x)$ tends to decay
exponentially from the boundary in the limit: $L\rightarrow \infty$
(dotted line). The decay length (the localization length) $\xi$ was
found to be the transport mean free path, which is
$21a$ ($50a$) for $\omega=1.65c/a$ ($0.72c/a$). We therefore rescale
$x$ into $x/\xi$ and present results for five different sample
lengths, $L/\xi= 2.5, 5, 10, 15 $ and $20$.
We see that for different frequencies,
the simulation results (squares and circles) overlap, signaling the
scaling behavior independent of the parameters of random media. It
is obvious that, except
near the boundaries, the results from simulation are significantly
larger than those from the SCLD model (dashed lines). The deviation
is prominent for large $L/\xi$\,, where the results from the SCLD
model converge to the sum of two truncated exponentials, decaying
from their respective boundaries. {\it Does localized waves in open
media display diffusive transport?} Our analytical prediction below
provides a definitive answer to this conceptually important
question. In particular, the analytical result of $D(x)/D_0$\,,
Eq.~(\ref{DSL}), is in excellent agreement with numerical
simulations (solid lines).

{\it Exact microscopic formalism.}---Referring to a separate
publication for technical details, we turn now to outline the proof
and the strategy is as follows. In the framework of the
supersymmetric field theory of localization \cite{Tian08,Efetov97},
we calculated explicitly the correlation function ${\cal Y}
(x,x')$\,, and found that it solves the local diffusion equation. In
doing so, we managed to calculate the weak localization correction,
$\delta D(x)$\,. Then, we found the Gell-Mann--Low equation of the
local diffusion coefficient, $D(x)=D_0+\delta D(x)$\,, which
eventually leads to Eq.~(\ref{DSL}).

In the present context the supersymmetric technique has many
advantages over others. The key ingredient is the introduction of a
``spin'' $Q$ (to be defined below) that encapsulates wave
interferences by fluctuations of the ``spin direction'': the larger
the fluctuation, the stronger the localization effect. With the help
of the $Q$-spin, the picture of localization in open media is
analogous to that of the more familiar problem---finite classical
ferromagnetic spin chains (but formal treatments are not). In the
latter system, the two end spins are fixed and parallel, and the
spin direction fluctuates elsewhere with a small (large) fluctuation
amplitude closed to the chain ends (midpoint). Translated to the
$Q$-spin language, such inhomogeneous fluctuations reflect the
spatial inhomogeneity of wave interferences in open media---the very
mechanism of local diffusion. Most importantly, the feature that the
two $Q$-spins at the boundaries are ``parallel'' takes all
disorder-induced resonant transmissions into full account. Thus, the
phenomenology of local diffusion is substantiated by a completely
microscopic formalism, albeit in an elegant manner.

Formally, the ``spin'' $Q$ is an $8\times 8$ supermatrix defined on
the advanced-retarded (ar), bosonic-fermonic (bf) and time-reversal
(tr) sector. The ``ar'' sector accommodates different analytic
structures of the advanced (retarded) Green function. The ``bf''
sector accommodates the supersymmetry: the diagonal (off-diagonal)
matrix elements are commuting (anti-commuting) numbers.  The ``tr''
sector accommodates the time-reversal symmetry. The leakage enters
through the boundary constraint: $Q(0)=Q(L)=\sigma_3^{\rm ar}
$\,. Here, $\sigma^X_3={\rm diag}(1,-1)$\,, $X={\rm ar, bf, tr}$\,.
Then, ${\cal Y}(x,x')$ is exactly expressed as
\begin{eqnarray}
{\cal Y}(x,x') = \frac{\pi \nu}{2^7} \int D[Q] e^{-F[Q]} {\rm str}
[\sigma_3^{\rm bf}(1+\sigma_3^{\rm ar})(1- \sigma_3^{\rm tr})
\nonumber\\
\times Q(x) (1-\sigma_3^{\rm ar})(1-\sigma_3^{\rm tr}) \sigma_3^{\rm
bf}Q(x')] \,. \label{DC}
\end{eqnarray}
The action, $F[Q]= -\frac{\pi\nu D_0}{8} \int_0^L dx\, {\rm str}\,
(\partial_x Q)^2$ ($\nu$ the density of states per unit length and
``str'' the supertrace), is the energy cost of the $Q$-field
fluctuations. It introduces a characteristic scale--the localization
length $\xi \propto \pi\nu D_0$\,.

Importantly, the mean field $Q(x)=\sigma_3^{\rm ar}
$\,, compatible with the boundary constraint, minimizes the action
and describes a vanishing wave intensity background across the
sample. Bearing this in mind, we introduced the
parametrization: $Q=(1+iW) \sigma_3^{\rm ar}
(1+iW)^{-1}$\,, where $W(x)$ anti-commutes with $\sigma_3^{\rm ar}$
and vanishes at $x=0,L$\,, and performed the $W$-expansion. By
keeping the leading order $W$-expansions, we obtained from
Eq.~(\ref{DC}) the bare correlation function $\mathscr{Y} (x,x')$\,,
which solves
$ -D_0 \partial_x^2 \mathscr{Y} (x,x') = \delta(x-x')$ with the
boundary condition: $\mathscr{Y} (0,x')= \mathscr{Y} (L,x')=0$\,.
Calculating the leading wave interference corrections to
$\mathscr{Y} (x,x')$\,, we found
\begin{eqnarray}
{\cal Y}(x,x') = \mathscr{Y} (x,x') - \!\! \int\!\! dy\, \mathscr{Y}
(x,y) \partial_y \delta D(y) \partial_y \mathscr{Y}(y,x')
\label{localdiffusion}
\end{eqnarray}
with the leading order weak localization correction read
\begin{equation}
\frac{\delta D(x)}{D_0}= \alpha \frac{\mathscr{Y} (x,x)}{\pi\nu}
-\frac{1}{2}(1-\alpha^2)\left[\frac{\mathscr{Y}
(x,x)}{\pi\nu}\right]^2 \,, \label{WL}
\end{equation}
where $\mathscr{Y} (x,x)=(L-x)x/(D_0 L)$\,, and $\alpha=-1$ for the
orthogonal symmetry while $\alpha=0$ for the unitary symmetry.
Eq.~(\ref{WL}) justifies that $D(x)$ depends on $x$ via $\lambda$\,.

For the unitary symmetry the first term of Eq.~(\ref{WL}) vanishes,
reflecting that the one-loop interference
is infeasible. Thus, the local diffusion and scaling behavior of
$D(x)$ are universal concepts, extrinsic to the time-reversal
symmetry that is required by the SCLD model.

{\it Scaling theory of local diffusion and disorder-induced resonant
transmission.}---We now make an important observation: ${\cal
G}(\lambda)= \nu D(x)/(\xi \lambda)$ and $\lambda$\,, formally, play
the role of the ``Thouless conductance'' \cite{Anderson79} and the
``system size'', respectively. Indeed, from Eq.~(\ref{WL}) we found
\begin{eqnarray}
\frac{d \ln {\cal G}}{d\ln \lambda} = \beta ({\cal G}) =
                             -1 + c_1 {\cal G}^{-1} + c_2 {\cal G}^{-2}+ \cdots ,\, {\cal
                             G}\gg 1\,,
\label{scaling}
\end{eqnarray}
with $c_1 < 0$ for $\alpha=-1$ and $c_1=0,\,c_2 < 0$ for
$\alpha=0$\,. This is fully analogous to the usual one-parameter
scaling theory of quasi-$1$D localization \cite{Anderson79} where,
in particular, the perturbative expansion of the $\beta$-function
also finds its origin in weak localization
\cite{Vollhardt80,Efetov97}.

From Eq.~(\ref{WL}), we further found that the weak localization
corrections of open and infinite media, including the coefficients,
are identical except that the returning probability, $\mathscr{Y}
(x,x)$\,, replaces that of infinite media. This duality persists in
all the higher-order weak localization corrections and, as such, the
$\beta$-function here is identical to that of the usual
one-parameter scaling theory \cite{Anderson79}. Identifying this
duality, we followed Refs.~\cite{Anderson79,Anderson80} to
extrapolate Eq.~(\ref{scaling}) into the regime of ${\cal G}\ll
1$\,, obtaining
\begin{eqnarray}
\beta({\cal G}) = \ln {\cal G} \,, \qquad {\cal G}\ll 1 \,.
\label{scaling1}
\end{eqnarray}

The scaling theory of local diffusion namely Eqs.~(\ref{scaling})
and (\ref{scaling1}) is far beyond the reach of earlier theoretical
studies \cite{Tian08,Skipetrov08} and has far-reaching consequences.
(It may be reproduced within the formalism of
Ref.~\cite{Brouwer08}.) In particular, Eq.~(\ref{scaling1}) gives
${\cal G}(\lambda)\propto e^{-\lambda}$ for $\lambda \rightarrow
\infty$ and thus Eq.~(\ref{DSL}) which, as shown below, fully
captures the rare disorder-induced resonant transmission. Solving
the static local diffusion equation, we found that the
ensemble-averaged transmission is $\langle T(L)\rangle \propto
[\int_0^L dx /D(x)]^{-1}$ (This is regardless of the explicit form
of $D(x)$\,, as first noticed in Ref.~\cite{Lagendijk00}.) Inserting
Eq.~(\ref{DSL}) into it gives $\langle T(L)\rangle \propto
e^{-L/(4\xi)}$ for $L\gg \xi$\,. On the other hand, noticing that
$T(L) = e^{-\gamma L}$\,, we found that the typical transmission
gives $d\langle \ln T(L)\rangle/dL = -\xi^{-1}$ from the Gaussian
distribution of $\gamma$ [cf. Eq.~(\ref{Lfluctuations})]. Thus, the
localization lengths obtained by the arithmetic and geometric means
differ by a factor $4$ irrespective of orthogonal or unitary
symmetry. This is in agreement with the result of the random matrix
theory, and is because that $\langle T(L)\rangle$ is dominated by
disorder-induced resonant transmissions \cite{Beenakker97}.

\begin{figure}
 \begin{center}
 \includegraphics[width=8.0cm]{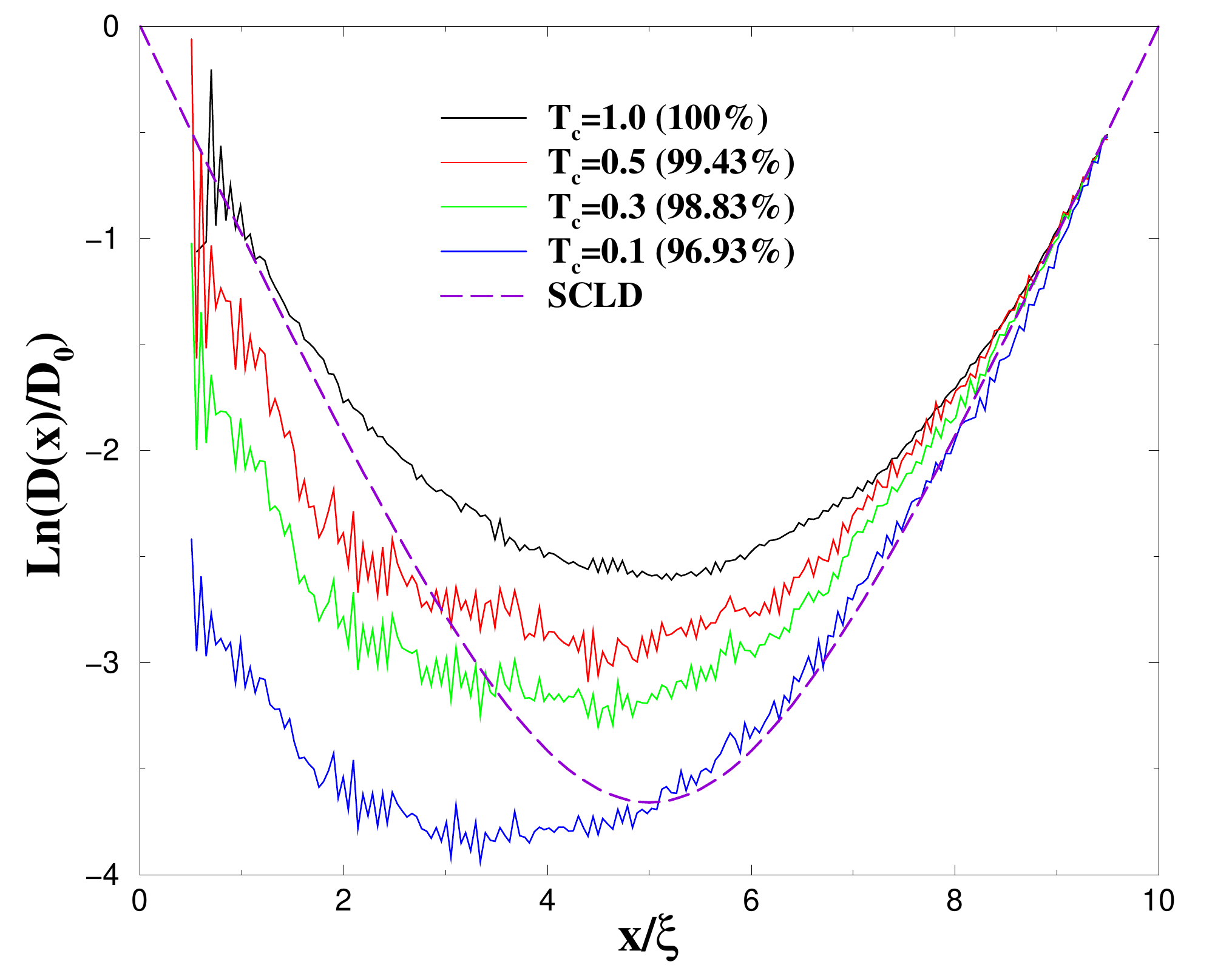}
\end{center}
 \caption{Simulations results obtained by eliminating the high-transmission ($T>T_c$)
 states in the original
 ensemble ($\omega=1.65c/a$ and $L/\xi=10$). The source is placed at $x=0$\,.}
 \label{Tsuppression}
\end{figure}

To further study effects of rare high-transmission states we
analyzed all the samples in Fig.~\ref{1Dlocal} ($\omega=1.65 c/a$).
First of all, the distribution of $\ln T$ is well fitted by the
normal distribution, with the average $\approx -L/\xi$ and the
variance $\approx 2L/\xi$\,. This confirms the Gaussian distribution
of $\gamma$\,. Then, we intentionally eliminated a small fraction of
high-transmission ($T>T_c$) states (composed mostly of singly
localized states and of a small portion of necklace states
\cite{Pendry94}), from the original ensemble ($L/\xi=10$) and
re-computed $D(x)/D_0$\,. As shown in Fig.~\ref{Tsuppression},
strikingly, even when the fraction of removed states is as small as
$0.6\%$ (solid line, in red), the simulation result deviates
drastically from the original one (solid line, in black) and is
asymmetric, signaling the breakdown of local diffusion. Thus, we
found that rare high-transmission states are essential to establish
local diffusion and scaling behavior.

{\it Conclusions.}---We found in quasi-$1$D localized samples a
scaling behavior of the (static) local diffusion coefficient
capturing all the rare disorder-induced resonant transmission. Our
findings show unambiguously that the prevailing SCLD model is valid
only if rare disorder-induced resonant transmissions are negligible
which, nevertheless, play a decisive role in transport of localized
waves.

The found phenomenon is intrinsic to finite-sized samples with open
boundaries and does not exist in an infinite sample. It is an
unconventional diffusion phenomenon in the sense that the diffusion
coefficient can drop by many orders of magnitude as the position
changes from the boundary to the midpoint. (For diffusive samples,
such a position-dependence is weak and thus does not lead to any
interesting phenomenon other than ordinary diffusion.) It is such a
drastic change (in the diffusion coefficient) that leads to a global
localization behavior as shown in the scaling of the average
transmission which decays exponentially with sample size.

Our theory has many immediate applications. For example, it can be
directly used to
study the speckle pattern of scattered waves which has recently
attracted considerable attentions. It may also be generalized to
higher dimensions for studying disorder-induced resonant
transmissions in (or close to) the localization regime. This issue
has practical applications such as
random laser.


We thank A. Z. Genack for discussions, and A. Altland, D. Basko, A.
A. Chabanov, M. Garst, A. Kamenev, and T. Nattermann for
conversations. Work supported by SFB TR12 of the DFG and the HK RGC
No. 604506.




\end{document}